**Innate immunity, assessed by plasma NO measurements, is not suppressed during the incubation fast in eiders.**

Sophie Bourgeon*[1], Thierry Raclot[1], Yvon Le Maho[1], Daniel Ricquier[2] and François Criscuolo[2]

[1] *IPHC, Département Ecologie, Physiologie et Ethologie, UMR 7178, 23 rue Becquerel,*

*F-67087 Strasbourg Cedex 2, France*

[2] *Université Paris V René Descartes, Site Necker enfants-Malades, CNRS-UPR 9078, 156 rue*

*de Vaugirard 75730 Paris Cedex 15, France*

Running title: Innate immunity in incubating eiders

*To whom correspondence should be addressed:

Sophie Bourgeon

IPHC, Département Ecologie, Physiologie et Ethologie, UMR 7178, 23 rue Becquerel,

F-67087 Strasbourg Cedex 2, France

Tel: (33) 3.88.10.69.15

Fax: (33) 3.88.10.69.06

e-mail:sophie.bourgeon@c-strasbourg.fr

**ABSTRACT: Innate immunity, assessed by plasma NO measurements, is not suppressed during the incubation fast in eiders.**

Immunity is hypothesized to share limited resources with other physiological functions and may mediate life history trade-offs, for example between reproduction and survival. However, vertebrate immune defence is a complex system that consists of three components. To date, no study has assessed all of these components for the same animal model and within a given situation. Previous studies have determined that the acquired immunity of common eiders (*Somateria mollissima*) is suppressed during incubation. The present paper aims to assess the innate immune response in fasting eiders in relation to their initial body condition. Innate immunity was assessed by measuring plasma nitric oxide levels (NO), prior to and after injection of lipopolysaccharides (LPS), a method which is easily applicable to many wild animals. Body condition index and corticosterone levels were subsequently determined as indicators of body condition and stress level prior to LPS injection. The innate immune response in eiders did not vary significantly throughout the incubation period. The innate immune response of eiders did not vary significantly in relation to their initial body condition but decreased significantly when corticosterone levels increased. However, NO levels after LPS injection were significantly and positively related to initial body condition, while there was a significant negative relationship with plasma corticosterone levels. Our study suggests that female eiders preserve an effective innate immune response during incubation and this response might be partially determined by the initial body condition.

 **INTRODUCTION**

3     Trade-offs between life history components are a central concept in evolutionary ecology.

4 The publication of a paper by Hamilton and Zuk in 1982 [1] generated a considerable interest

5 in the role that parasites might play in the evolution of reproductive strategies. This stimulated

6 the emergence of the field of ecological immunology, which investigates the role of immune

7 effector systems in determining host fitness in the wild [2]. The observation that a heavy

8 parasite load during reproduction is associated with a reduced immune response [3] and a

9 decreased adult survival rate [4], led to the hypothesis that immunity must share limited

10 resources with other physiological functions and this may to some extent underpin the costs of

11 reproduction [5].

12     Recent experiments have indicated that high reproductive effort increases metabolic

13 rate, leading to oxidative stress [6,7] due to increased generation of oxidative metabolites and

14 free radicals [8]. Immunosuppression during reproduction has been viewed as an undesirable

15 consequence of a high metabolic rate [9]. However, even if a high metabolic rate during

16 reproduction might contribute to immunosuppression, it is most likely not the only factor

17 involved [9]. As recently outlined by Viney et al. [10], a "maximum immune response" does

18 not necessarily mean that the response is "optimal", since an immunosuppression might be of

19 adaptive value [9] and then accepted as a cost by the organism. Firstly, since an activated

20 immunity enhances oxidative stress [8], it carries a non-negligible risk of autoimmune

21 pathology [11,12]. Immunosuppression could therefore be explained by the

22 immunopathology-avoidance hypothesis [9]. In this context, the control of the production of

23 nitric oxide (NO) by immune cells is primordial since overproduction of NO is involved in

24 sepsis shock [13]. Secondly, the resource-limitation hypothesis suggests that different

25 functions compete for limited resources and that investment in costly behaviours, such as reproduction, reduces the amount of resources available to immune defense [9]. This second hypothesis assumes that there is an energetic or nutritional cost associated with the immune system [9,14,15]. However, evidence for an energetically costly immune response is still equivocal [9].

Vertebrate immune defense is one of the most complex biological phenomena. It consists of three components (innate immunity, acquired humoral immunity, and acquired cellular immunity) between which trade-offs may appear [5,10,16-18]. Consequently, multiple immune assays, which challenge different components of the immune system, should be used to gain a better understanding of overall immunocompetence [19]. Since assessment of acquired immunity has proven insufficient for ecological immune studies [16,20], assaying the innate component is of primary importance.

Eider ducks are a useful model for studying the intensity of the immune response in relation to body condition, since females rely solely on their stored body fat during incubation [21]. Previous studies have shown that the acquired immunity of common eider ducks (*Somateria mollissima*) is suppressed during the incubation fast [18,22], and that its experimental activation has strong negative effects on the fitness of female eiders [22]. In fact, immunocompetence may be fixed by fat stores, the main energetic reserves of the organism that determine individual quality of reproductive adults [5,23]. Thus, the decrease in acquired immune response of female eiders can be understood as a consequence of fat stores depletion. However, whether and to what extent eider's whole immune response during reproduction depends on individual quality and is suppressed when body conditions deteriorate is still under debate [24].

Based on these data, we tested whether during the incubation fast, female eiders preserve or not an impenetrable first line of defense. In other words, do eiders maintain a highly effective innate immune response to compensate for the loss of their acquired immune components?

To date, only one study has assessed the constitutive innate immunity by using an *in vitro* assay [25]. We used here an alternative method, commonly used in immunology studies, which is easily applicable to animals in the wild. The innate immune response consists of the activation of phagocyte cells as heterophils and macrophages [26]. Destruction of recognized pathogens by those cells is mainly driven by the production of NO [27], which can be experimentally stimulated by injection of lipopolysaccharides (LPS) [28]. We measured plasma NO concentrations in breeding female eider ducks before and after LPS injection at various stages during their incubation period, while also considering their initial body condition. We assessed body condition by measuring corticosterone levels, an increase in one component being strongly associated with a decline in the other [29,30], and also by calculating a body mass index (mass/tarsus length$^3$).

# MATERIAL AND METHODS

The study was conducted in a common eider colony on Storholmen Island, Kongsfjorden, Svalbard Archipelago (78°55' N, 20°07' E) from June to July 2005. The breeding population of eiders in the study site consisted of about 900 nests. Females in the study area laid between 1 and 6 eggs, but a clutch size of 4 to 5 eggs was most common (49.35 % and 24.62 % respectively, N=845). Eider ducklings are precocial and are cared for by the female only. Incubation lasts between 24 and 26 days [21]. All birds started their incubation between June 07 and June 13, the main laying period for the colony. Ducks that laid their eggs after this period were not considered in this study. Ambient temperatures in June and July ranged from 2 to 10°C.

Groups and sampling protocol

Nests were checked at least every two days throughout the study period. This was done to determine initial clutch size but also to investigate the rate of egg predation and nest desertion. A clutch of eggs was considered complete when no additional egg was laid during a two-day period [31]. Female eiders were caught on their nests using a bamboo pole with a nylon snare. Blood (1- 2 ml) was collected from the brachial vein within three minutes of capture. It was stored in tubes containing EDTA (an anticoagulant agent) and kept on ice until being centrifuged at 10, 000 rpm for five minutes at 4°C in the laboratory. Plasma samples and blood cells were separately stored at −20°C and plasma was subsequently used to measure corticosterone and nitric oxide levels. After blood sampling, body size was recorded (wing and tarsus lengths) and birds were weighed with a portable electronic balance (± 2 g). Initial body mass (after laying the last egg) was extrapolated from body mass at sampling as described by Criscuolo et al. [32]. Finally, to measure the innate immune response, each female was injected with a solution of lipopolysaccharides (LPS) intra-peritoneally (see below). To assess female body condition, the following index was used: female initial body mass was divided by (tarsus length)$^3$, as reported by MacColl and Hatchwela [33]. Mean body condition of 51 females was $8.80 \pm 0.10$ (mean $\pm$ SE, N=51). Consequently, we classified females with an index $\leq 8.80$ as being in bad condition, while females with an index $> 8.80$ as being in good condition.

A total of 51 females with clutch sizes ranging from 3 to 6 eggs ($4.06 \pm 0.09$ eggs, N=51) (mean $\pm$ SE) was used in this study. All birds were caught twice. To prevent nest desertion we only caught birds that were already incubating for at least five days. To cover the entire incubation period we caught individuals that were at different stages of their incubation period ($17.94 \pm 0.77$ days into incubation, N=51) (mean $\pm$ SE).

Innate immune response: LPS injection

To evaluate the innate immune response, we first collected blood samples from incubating females and subsequently injected $0.35$ mg.kg$^{-1}$ LPS intra-peritoneally (Sigma L-2630, diluted in phosphate-buffered saline). Birds were recaught 24 h after injection and another blood sample was taken. Nitric oxide levels were subsequently determined (see below) prior to and 24 h after LPS injection. The 'innate immune response' was calculated as the increase (in percentage) in nitric oxide production following LPS injection.

Assessment of the nitric oxide levels

There are various assays for the quantification of NO in biological models as reviewed by Archer [34]. The free radical NO has a short half-life *in vivo* since it is quickly oxidized, mainly to the inorganic nitrite ($NO_2^-$) and nitrate ($NO_3^-$). Diazotization is a standard technique for indirect quantification of NO production. It has been previously used in chickens [35] and is based on the reaction of $NO_2^-$ with Griess components (sulphanilamide and N-(1-naphthyl)ethylene-diamine, Sigma G4410-10G), and the generation of a purple-azo-dye product, which is measured with a spectrophotometer at 550 nm [35]. This method requires previous enzymatic reduction of $NO_3^-$ to $NO_2^-$ using bacterial nitrate reductase (Sigma N-7265).

In our assessment, plasma (10 μl) was diluted 1:2 with sterile distilled water (10 μl) and incubated for two hours at ambient temperature with 80 μl of a medium composed of nitrate reductase (Sigma N-7265, 0.4 u.mL$^{-1}$), β-NADPH (Nicotinamide adenine di-nucleotide phosphate, Sigma 7505, 1 mg.mL$^{-1}$) and HEPES (GIBCO 15630-056, 0.1 M). A volume of 100 μL of Griess reagent was added to each sample and the optical density was measured at 550 nm after 15 minutes. Nitrite levels were determined with a standard curve generated from a $NaNO_2$ solution at 2.5 mM. Sodium EDTA was used as anticoagulant and no ultrafiltration of plasma was done, as previously recommended, to avoid interference in $NO_3^-$ quantification [36]. The absorbance of background coloured plasma was controlled for the non-specific increase in absorbance after the Griess reaction due to hemolysis or to the presence of blood haemoproteins [36]. No particular elevated backgrounds due to haemolysis were found in the samples.

### Assessment of the corticosterone levels

Corticosterone concentration was determined by radioimmunoassay (RIA) in our laboratory using an [125]I RIA double antibody kit from ICN Biomedicals (Costa Mesa, CA, USA).The corticosterone RIA had an intraassay variability of 7.1 % (N=10 duplicates) and an interassay variability of 6.5 % (N=15 duplicates).

1    <u>Statistical analyses</u>

2    Statistical analysis was conducted with SPSS 12.0.1 (SPSS Inc., Chicago, Illinois, USA).

3    Values are means ± standard error (SE). When data were not normally distributed

4    (Kolmogorov-Smirnov test, $P<0.05$), values were log transformed before using parametric

5    tests. When transformations were not possible, non parametric tests were used.

1  **RESULTS**

3  Table 1 provides details of the female eiders used in this study. Initial body mass, initial

4  clutch size, and wing length were not significantly different between the two groups (good

5  and bad body condition). Body mass (1600 ± 23 g, N=51) was inversely related to sampling

6  date during the fast (Spearman rank correlation: $r_s$=-0.68, N=51, P<0.0001). There was no

7  significant relationship between female initial body mass and initial clutch size (Spearman

8  rank correlation: $r_s$=-0.03, N=51, P=0.81).

10  <u>Nitric oxide production in relation to incubation length</u>

11  As expected, LPS injection stimulated NO production. Indeed, NO levels were significantly

12  increased by 70 % following LPS injection (repeated measures ANOVA: $F_{1,50}$=107.02,

13  P<0.0001) (106.82 ± 7.85 µM, N=51 before LPS injection versus 179.13 ± 9.64 µM, N=51

14  after LPS injection).

15  The percentage of increase in NO production following LPS injection (the 'innate immune

16  response') did not vary significantly throughout the incubation period, ranging from 110 to 74

17  % (linear regression: $r^2$=0.07, N=51, P=0.06). By contrast, NO levels before LPS injection

18  (NO1) increased significantly by about 40 % as the incubation progressed (range: 86.66-

19  120.94 µM; linear regression: $r^2$=0.12, N=51, P=0.01) (Figure 1). However, NO levels after

20  LPS injection (NO2) did not vary significantly throughout the incubation period (range:

21  168.30-186.72 µM; linear regression: $r^2$=0.04, N=51, P=0.15). Finally, there was a significant

22  negative relationship between the innate immune response and NO1 levels (linear regression:

23  $r^2$=0.57, N=51, P<0.0001) but not NO2 levels (linear regression: $r^2$=0.03, N=51, P=0.23)

24  indicating that when NO levels were high before LPS injection, no further increase was

25  observed after LPS injection (Figure 2).

1   <u>Nitric oxide production in relation to female body condition</u>

2   We first assessed female body condition by dividing their initial body mass by their (tarsus

3   length)$^3$. The body condition index was not significantly related to neither innate immune

4   response nor NO1 levels (linear regression: r²=0.00, N=51, P=0.99; r²=0.04, N=51, P=0.18,

5   respectively). By contrast, there was a significant positive relationship between body

6   condition index and NO2 levels (linear regression: r²=0.12, N=51, P=0.01), so that females in

7   good body condition had greater NO2 values than females in bad body condition (Figure 3).

8   We subsequently assessed female body condition by measuring their corticosterone levels.

9   Plasma corticosterone levels did not vary significantly over the incubation period (linear

10  regression: r²=0.001, N=51, P=0.82). Mean corticosterone level was 28.70 ± 4.66 ng.ml$^{-1}$

11  (range: 22-33 ng.ml$^{-1}$; N=51). We found a significant negative relationship between body

12  condition index and plasma corticosterone levels (linear regression: r²=0.12, N=51, P=0.01).

13  Hence, females in good body condition had lower levels of corticosterone than females in bad

14  body condition.

15  There was a significant negative relationship between plasma corticosterone levels and the

16  innate immune response (% increase of NO after LPS injection) (linear regression: r²=0.10,

17  N=51, P=0.02) (Figure 4), whereby high levels of corticosterone were associated with a low

18  innate immune response. We furthermore found a significant negative relationship between

19  plasma corticosterone and NO2 levels (linear regression: r²=0.11, N=51, P=0.01) (Figure 5),

20  where high levels of corticosterone were associated with low values of NO2. However, there

21  was no significant relationship between plasma corticosterone and NO1 levels (linear

22  regressions: r²=0.04, N=51, P=0.15).

# DISCUSSION

Our study is the third part of an investigation into the immunocompetence of female eider ducks during the incubation fast, and allows us to give an overview of how the different components of immunity are affected during this phase. Suppression of both T-cell-mediated and humoral acquired immune responses were previously observed in incubating female eiders [18,22,37], but no significant correlation between these two components has been observed [18]. This immunosuppression appears to be adaptive since it enhances the future survival of the females [22]. Roitt et al. [38] suggested that, birds should consequently keep a high innate barrier to compensate for the loss of the acquired immunity, so that pathogens can be eliminated before the acquired components are triggered. Support for this hypothesis comes from indirect measurements of the innate immunity in eiders (i.e. heterophil/lymphocyte counts) by Hanssen et al. [39]. Furthermore, the plasma NO production measured in the present study clearly illustrates that the innate response in our eiders was not suppressed during fasting. Similarly, studies on severely starved chickens showed an enhanced innate immunity while the lymphocyte response (as part of the acquired immune response) was suppressed [40]. A possible explanation for these observations is that the recruitment of phagocyte cells from the innate system is less energy-consuming than the activation of humoral and cellular acquired immunities. Favouring the innate system could also act as a buffer against the fitness costs associated with acquired immunities [22]. However, it is still unclear if the activation of an immune response is energetically costly and evidence from a number of immuno-ecological studies is equivocal. Demas et al. [41] showed that oxygen consumption of mice was increased by 15-20 % after an immunoglobulin response was triggered. Svensson et al. [42], however, found only a weak increase in the BMR of blue tits (*Parus caeruleus*) after antibody response challenge. Similarly, Verhulst et al. [15] reported that the metabolic costs of eliciting a humoral immune response are negligible in zebra finches (*Taeniopygia guttata*). Also, Hanssen et al. [22] did not find an accelerated body mass loss in female eiders responding to an injection of antigens. Therefore, the nature of the costs linking immune responses to reproductive effort and overall energy metabolism still escapes our understanding. However, since apoptosis of leukocytes is naturally induced by free radicals [43], oxidative damage could mediate such a cost [8,9]. Finally, despite the fact that innate immunity was not suppressed during the incubation fast in eider ducks, it nonetheless seemed to be limited to a certain threshold. In fact, while NO1 levels significantly increased as incubation progressed, NO2 levels did not vary significantly over the course of the incubation period. As a consequence, the innate immune response (NO2-NO1/NO1) did not vary throughout incubation. Moreover, we found a negative relationship between NO1 levels and the innate immune response, so that higher NO1 levels were associated with a lower innate immune response. This could be a potential mechanism to reduce the risks of oxidative damage, since this non-specific defense leads to an over-production of free radicals. NO is a potent vasodilator playing a crucial role in sepsis shock [13], and its production must be then carefully controlled. Moreover, NO is produced not only by macrophages but also by endothelial cells and, interestingly, by adipose tissue [44], which may explain the unexpected variations of NO1 levels (before LPS injection and immune system challenge) in eiders. In this context, it would be interesting to test both the immunosuppressive effect of an increased oxidative stress but also changes in basal NO levels during the incubation fast in future experiments.

Trade-offs between different components of the immune system were reported in red jungle fowl (*Gallus gallus*) [17]. Before the breeding season cell-mediated immunity and the proportion of lymphocytes of male jungle fowl were positively correlated with comb length. By contrast, during the breeding season, males with large combs had lower levels of lymphocytes but a greater cell-mediated immunity [17]. The underlying mechanisms are still unknown. In a previous study on incubating eiders, humoral immunosuppression was correlated with an increase in plasma concentration of two heat shock proteins (HSP60 and HSP70) [12]. HSP's are known to be immunogenic for innate responses [45] and to promote lysis activity of the innate component members such as natural killer cells (NK) [46]. Hence, they might contribute to the maintenance of the innate immunity during fasting. NK cells are specialized in the destruction of infected host cells, which could partly compensate for lymphocyte deficiency against viral infections [26]. In addition, in previous studies, suppression of the acquired immunity was stronger in female eiders with a low initial body condition [39], while adverse effects on fitness from parasites were also more pronounced in these individuals [47]. We found that $NO_2$ levels in our eiders were positively correlated with initial body condition. Hence, the immunocompetence of eider ducks might to some degree depend on initial body condition [5]. This idea is supported by the fact that humoral immunity in mammals is controlled by body fat stores [23], which raises the question of the hormonal control of the immune system [48]. The peptide hormone leptin is secreted primarily by adipose tissue and has been shown to enhance a variety of immunological parameters in mammals [49,50] and birds [51]. Since circulating levels of leptin are generally proportional to the amount of body fat [52], decreases in body fat stores may affect immunity via changes in endocrine signaling with the immune system [48]. To examine the role of leptin for the immune function of fasting eiders, it would be useful to manipulate leptin concentrations [51] at different incubation stages. Corticosterone might also be a potential functional link between body condition and immune response. It was namely implicated in the *in vitro* inhibition of chicken lymphocyte proliferation [53] and in decreased phagocyte ability in starved rats [54]. Conversely, corticosterone can also promote phagocytosis [55] and NO production in synergy with prolactin [56]. This latter peptide has shown to be involved in incubation behaviour in eiders [57]. In this study, we reported that corticosterone, the primary avian stress hormone, tended to be negatively correlated with the innate immune response, so that high levels of corticosterone were associated with a low innate immune response.

In conclusion, we propose that in contrast to the fate of both acquired immune components, the innate immunity is not suppressed during the incubation fast of female eiders. Moreover, the levels of NO2 seem to depend on the initial amount of body reserves. Females in good body condition showed higher NO2 levels than birds in bad body condition. Similarly, females with high corticosterone values had low NO2 values and a decreased innate immune response. Nevertheless, the evolutionary significance of our observations remain to be validated. Finally, despite the fact that hormonal and neuroendocrine systems are important promoters of innate immunity [58], their interactions need to be investigated in further experiments.

1 **ACKNOWLEDGEMENTS**

3 We wish to thank Yalin Emre and Dr Sophie Rousset for their advice and help during the NO

4 assessment. This research was funded by The French Polar Institute Paul Emile Victor and the

5 Comité Interprofessionnel de la Dinde Française (CIDEF). The experiments were conducted

6 with the authorization of the French and Norwegian Ethics Committees and the Governor of

7 Svalbard.

1    **REFERENCES**

3    [1] Hamilton WD, Zuk M. Heritable true fitness and bright birds: a role for parasites? Science

4    1982; 218: 384-387.

5    [2] Sheldon BC, Verhulst S. Ecological immunology: costly parasite defences and trade-offs

6    in evolutionary ecology. Trends Ecol Evol 1996; 11: 317-321.

7    [3] Nordling D, Andersson M, Zohari S, Gustafsson L. Reproductive effort reduces specific

8    immune response and parasite resistance. Proc R Soc Lond B 1998; 265: 1291-1298.

9    [4] Ricklefs RE. Density dependence, evolutionary optimization, and the diversification of

10   avian life histories. Condor 2000; 102: 9-22.

11   [5] Norris K, Evans MR. Ecological immunology: life history trade-offs and immune defense

12   in birds. Behav Ecol 2000; 11: 19-26.

13   [6] Nilsson J-Å. Metabolic consequences of hard work. Proc R Soc Lond B 2002; 269: 1735-

14   1739.

15   [7] Wiersma P, Selman C, Speakman JR, Verhulst S. Birds sacrifice oxidative protection for

16   reproduction. Proc R Soc Lond B 2004; 271 (S5): S360-S363.

17   [8] von Schantz T, Bensch S, Grahn M, Hasselqist D, Wittzell H. Good genes, oxidative

18   stress and condition-dependent sexual signals. Proc R Soc Lond B 1999; 266: 1-12.

19   [9] Råberg L, Grahn M, Hasselquist D, Svensson E. On the adaptive significance of stress-

20   induced immunosuppression. Proc R Soc Lond B 1998; 265: 1637-1641.

21   [10] Viney ME, Riley EM, Buchanan KL. Optimal immune responses: immunocompetence

22   revisited. Trends Ecol Evol 2005; 20 (12): 665-669.

23   [11] MacDonald J, Galley HF, Webster NR. Oxidative stress and gene expression in sepsis.

24   Br J Anaesth 2003; 90: 221-232.

[12] Bourgeon S, Martinez J, Criscuolo F, Le Maho Y, Raclot T. Fasting-induced changes of immunological and stress indicators in breeding female Eiders. Gen Comp Endocrinol 2006; 147 (3): 336-342.

[13] Szabo C, Mitchell JA, Thiemermann C, Vane JR. Nitric oxide-mediated hyporeactivity to noradrenaline precedes the induction of nitric oxide synthase in endotoxin shock. Br J Pharmacol 1993; 108: 786-789.

[14] Martin II LB, Scheuerlein A, Wikelski M. Immune activity elevates energy expenditure of house sparrows: a link between direct and indirect costs? Proc R Soc Lond B 2002; 270: 153-158.

[15] Verhulst S, Riedstra B, Wiersma P. Brood size and immunity costs in zebra finches *Taeniopygia guttata*. J Avian Biol 2005; 36: 22-30.

[16] Gross WG, Siegel PB, Hall W, Domermuth CH, DuBoise RT. Production and persistence of antibodies in chicken to sheep erythrocytes. 2. Resistance to infectious diseases. Poult Sci 1980; 59: 205-210.

[17] Zuk M, Johnsen TS. Seasonal changes in the relationship between ornamentation and immune response in red jungle fowl. Proc R Soc Lond B 1998; 265: 1631-1635.

[18] Bourgeon S, Criscuolo F, Le Maho Y, Raclot T. Phytohemagglutinin response and immunoglobulin index decrease during incubation fasting in female Common Eiders. Physiol Biochem Zool 2006; 79 (4): 793-800.

[19] Adamo SA. How should behavioural ecologists interpret measurements of immunity? Anim Behav 2004; 68: 1443-1449.

[20] Keil D, Luebke R, Pruett S. Quantifying the relationship between multiple immunological parameters and host resistance: probing the limits of reductionism. J Immunol 2001; 167: 4543-4552.

[21] Korschgen CE. Breeding stress of female Eiders in Maine. J Wild Manage 1977; 41 (3): 360-373.

[22] Hanssen SA, Hasselquist D, Folstad I, Erikstad KE. Costs of immunity: immune responsiveness reduces survival in a vertebrate. Proc R Soc Lond B 2004; 271: 925-930.

[23] Demas GE, Drazen DL, Nelson RJ. Reductions in total body fat decrease humoral immunity. Proc R Soc Lond B 2003; 270: 905-911.

[24] Eraud C, Duriez, O, Chastel, O, Faivre B. The energetic cost of humoral immunity in the Collared Dove, *Streptopelia decaocto*: is the magnitude sufficient to force energy-based trade-offs? Funct Ecol 2005; 19: 110-118.

[25] Tieleman BI, Williams JB, Ricklefs RE, Klasing KC. Constitutive innate immunity is a component of the pace-of-life syndrome in tropical birds. Proc R Soc Lond B 2005; 272: 1715-1720.

[26] Tosi MF. Innate immune responses to infection. J Allergy Clin Immunol 2005; 116: 241-249.

[27] Crippen TL, Sheffield CL, Haiqi H, Lowry VK, Kogut MH . Differential nitric oxide production by chicken immune cells. Dev Comp Immunol 2003; 27: 603-610.

[28] Lillehoj HS, Li G. Nitric oxide production by macrophages stimulated with coccidia sporozoites, lipopolysaccharide, or interferon-γ, and its dynamic changes in SC and TK strains of chickens infected with *Eimeria tenella*. Avian Dis 2004; 48: 244-253.

[29] Harvey S, Phillips JG, Rees A, Hall TR. Stress and adrenal function. J Exp Zool 1984; 232: 633–645.

[30] Love OP, Chin EH, Wynne-Edwards KE, Williams TD. Stress hormones: a link between maternal condition and sex-biased reproductive investment. Am Nat 2005; 166 (6): 751-766.

[31] Yoccoz NG, Erikstad KE, Bustnes JO, Hanssen SA, Tveraa T. Cost of reproduction in common eiders (*Somateria mollissima*): an assessment of relationships between reproductive

effort and future survival and reproduction based on observational and experimental studies. J Appl Stat 2002; 29: 57-64.

[32] Criscuolo F, Gabrielsen GW, Gendner JP, Le Maho Y . Recess behaviour and body mass regulation during extended incubation in Common Eider (*Somateria mollissima*). J Avian Biol 2002; 33: 83-88.

[33] MacColl ADC, Hatchwela BJ. Heritability of parental effort in a passerine bird. Evolution 2003; 57 (9): 2191-2195.

[34] Archer S. Measurement of nitric oxide in biological models. FASEB J 1993; 7: 349-360.

[35] Jarosinski KW, Ynis R, O'Connell PH, Markowski-Grimsrud CJ, Schat KA. Influence of genetic resistance of the chicken and virulence of Marek's disease virus (MDV) on nitric oxide responses after MDV infection. Avian Dis 2002; 46: 636-649.

[36] Ricart-Jané D, Llobera M, Lopez-Tejero MD. Anticoagulants and other preanalytical factors interfere in plasma nitrate/nitrite quantification by the Griess method. Nitric Oxide 2002; 6: 178-185.

[37] Hanssen SA, Hasselquist D, Folstad I, Erikstad KE. Cost of reproduction in a long-lived bird: incubation effort reduces immune function and future reproduction. Proc R Soc Lond B 2005; 272: 1039-1046.

[38] Roitt IM, Brostoff J, Male DK. Immunology. London, Mosby, 1998.

[39] Hanssen SA, Folstad I, Erikstad KE. Reduced immunocompetence and cost of reproduction in common eiders. Oecologia 2003; 136: 457-464.

[40] Hangalapura BN, Nieuwland MGB, De Vries Reilingh G, Buyse J, Van Den Brand H, Kemp B, Parmentier HK. Severe feed restriction enhances innate immunity but suppresses cellular immunity in chicken lines divergently selected for antibody responses. Poult Sci 2005; 84: 1520-1529.

1   [41] Demas GE, Chefer V, Talan MI, Nelson RJ. Metabolic costs of mounting an antigen-

2   stimulated immune response in adult and aged C57BL/6J mice. Am J Physiol 1997; 273:

3   R1631-R1637.

4   [42] Svensson E, Råberg L, Koch C, Hasselquist D. Energetic stress, immunosuppression and

5   the costs of an antibody response. Funct Ecol 1998; 12: 912-919.

6   [43] Guzik TJ, Korbut R, Adamek-Guzik T. Nitric oxide and superoxide in inflammation and

7   immune regulation. J Physiol Pharmacol 2003; 54: 469-487.

8   [44] Elizalde M, Ryden M, van Harmelen V, Eneroth P, Gyllenhammar H, Holm C, Ramel S,

9   Olund A, Arner P, Andersson K. Expression of nitric oxide synthases in subcutaneous adipose

10  tissue of nonobese and obese humans. J Lipid Res 2000; 41: 1244-1251.

11  [45] Gullo CA, Teoh G. Heat shock proteins: to present or not, that is the question. Immunol

12  Lett 2004; 94: 1-10.

13  [46] Hickman-Miller HD, Hildebrand WH. The immune response under stress: the role of

14  HSP-derived peptides. Trends Ecol Evol 2004; 25: 427-433.

15  [47] Hanssen SA, Folstad I, Erikstad KE, Oksanen A. Costs of parasites in common eiders:

16  effects of antiparasite treatment. Oikos 2003; 100: 105-111.

17  [48] Demas GE, Sakaria S. Leptin regulates energetic tradeoffs between body fat and

18  humoural immunity. Proc R Soc Lond B 2005; 272: 1845-1850.

19  [49] Lord GM, Matarese G, Howard JK, Baker RJ, Bloom SR, Lechler RI. Leptin modulates

20  the T-cell immune response and reverses starvation-induced immunosuppression. Nature

21  1998; 394: 897-901.

22  [50] Faggioni R, Feingold KR, Grunfeld C. Leptin regulation of the immune response and the

23  immuno-defiency of malnutrition. FASEB J 2001; 15: 2565-2571.

24  [51] Lõhmus M, Olin M, Sundström LF, Troedsson MHT, Molitor TW, El Halawani M.

25  Leptin increases T-cell immune response in birds. Gen Comp Endocrinol 2004; 139: 245-250.

[52] Lõhmus M, Sundström LF. Leptin and social environment influence the risk-taking and feeding behaviour of Asian blue quail. Anim Behav 2004; 68: 607-612.

[53] Hangalapura BN, Nieuwland MG, Buyse J, Kemp B, Parmentier HK. Effect of duration of cold stress on plasma adrenal and thyroid hormone levels and immune responses in chicken lines divergently selected for antibody responses. Poult Sci 2004; 83: 1644-1649.

[54] Salman H, Bergman M, Bessler H, Alexandrova S, Straussberg R, Zahavi I, Djaldetti M. Effect of three days starvation on the phagocytic activity of rat peritoneal macrophages. Acta Haematol 1998; 100 (1): 17-21.

[55] Rodriguez AB, Terron MP, Duran J, Ortega E, Barriga C. Physiological concentrations of melatonin and corticosterone affect phagocytosis and oxidative metabolism of ring dove heterophils. J Pineal Res 2001; 31: 31-38.

[56] Lourenco GA, Dorce VA, Paermo-Neto J. Haloperidol treatments increased macrophage activity in male and female rats: influence of corticosterone and prolactin serum levels. Eur Neuropsychopharmacol 2005; 15: 271-277.

[57] Criscuolo F, Chastel O, Gabrielsen GW, Le Maho Y. Circulating prolactin in a capital incubator : the Common Eider *Somateria mollissima.* Gen Comp Endocrinol 2002; 125: 399-409.

[58] Berczi I, Bertok L, Chow DA. Natural immunity and neuroimmune host defense. Ann NY Acad Sci 2000; 917: 248-257.

 **FIGURE LEGENDS**

3 **Figure 1.** Nitric oxide levels before LPS injection (NO1) in female eiders versus incubation

4      stage and according to their initial body condition: good (●) or bad (▽). LPS,

5      lipopolysaccharides.

7    **Figure 2.** Relationship between nitric oxide levels before LPS injection (NO1) and innate

8      immune response (%) in incubating female eiders according to their initial body condition:

9      good (●) or bad (▽).

11 **Figure 3.** Relationship between body condition index (mass/tarsus length$^3$) and nitric oxide

12      levels after LPS injection (NO2) in incubating female eiders. Females having an index $\leq 8.80$

13      or $> 8.80$ were classified as in bad (▽) or good (●) initial body condition, respectively. LPS,

14      lipopolysaccharides.

16 **Figure 4.** Relationship between corticosterone levels and innate immune response (%) in

17      incubating female eiders according to their initial body condition: good (●) or bad (▽).

19 **Figure 5.** Relationship between corticosterone levels and nitric oxide levels after LPS

20      injection (NO2) in incubating female eiders according to their initial body condition: good

21      (●) or bad (▽). LPS, lipopolysaccharides.

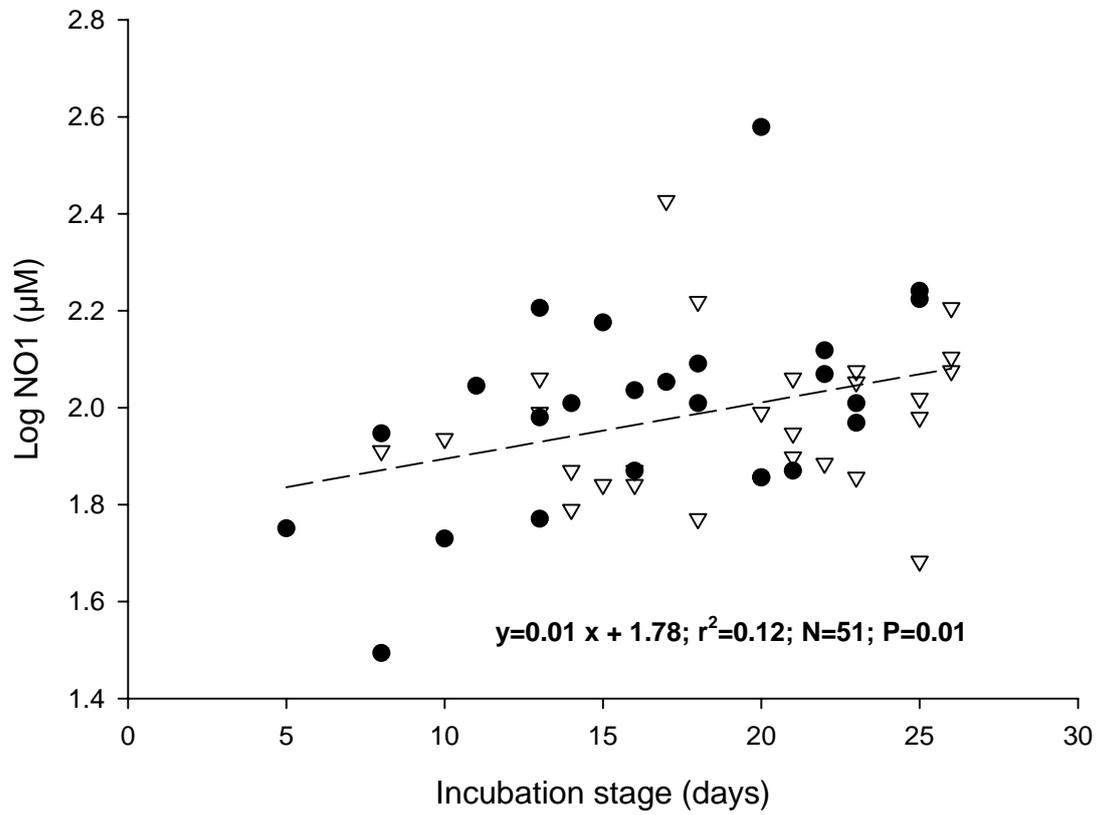

1 **FIGURE 2.**

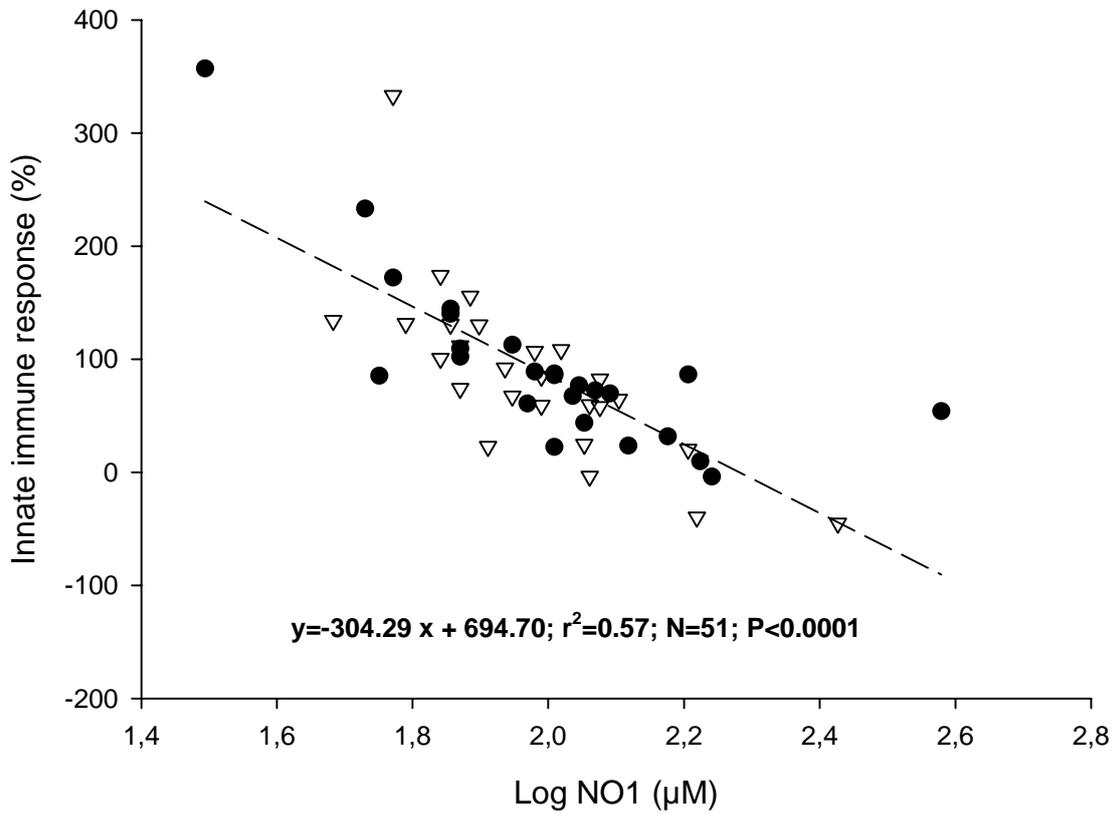

Innate immune response (%) vs Log NO1 (μM)

$y = -304.29 x + 694.70; r^2 = 0.57; N = 51; P < 0.0001$

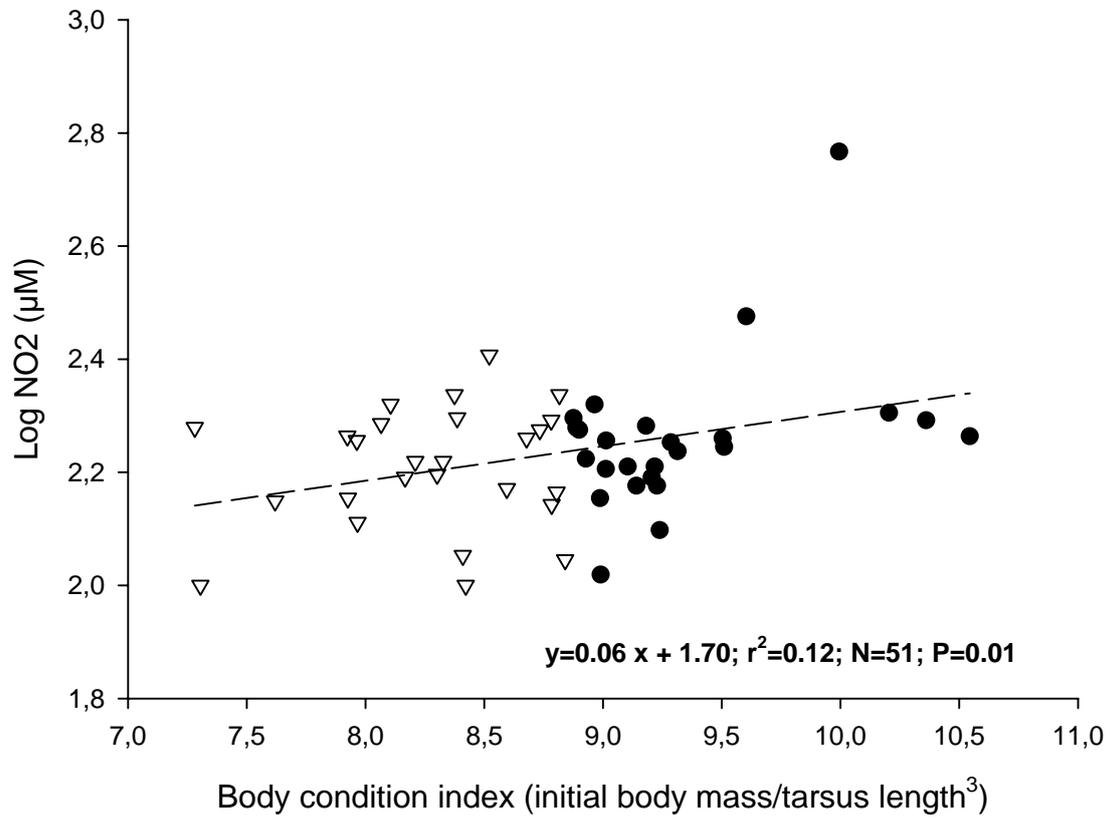

Body condition index (initial body mass/tarsus length$^3$)

y=0.06 x + 1.70; r$^2$=0.12; N=51; P=0.01

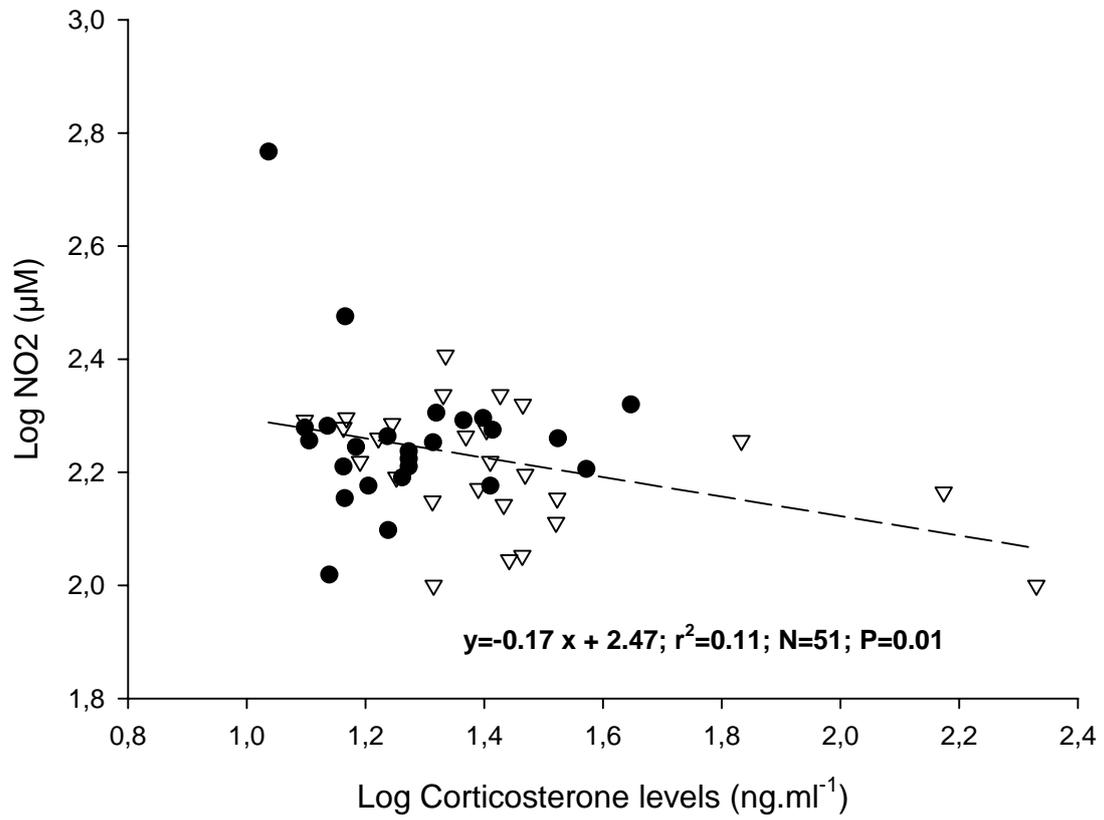

$$y=-0.17\ x + 2.47;\ r^2=0.11;\ N=51;\ P=0.01$$

3   **Table 1.** Profiles of female eiders according to their body condition. Body condition was

4   calculated by dividing female initial body mass by (tarsus length)$^3$. Females having an index

5   $\leq 8.80$ or $> 8.80$ were classified as in bad or good body condition, respectively. Values are

6   means $\pm$ SE.

| | Group 1: females in good body condition N=25 | Group 2: females in bad body condition N=26 | U-test | P |
|---|---|---|---|---|
| **Initial body mass (g)** | 2018 ± 23 | 1990 ± 24 | -0.48 | 0.63 |
| **Initial clutch size (eggs)** | 4.00 ± 0.13 | 4.11 ± 0.13 | -0.50 | 0.62 |
| **Tarsus length (cm)** | 6.00 ± 0.03 | 6.22 ± 0.02 | -4.48 | <0.0001 |
| **Wing length (cm)** | 29.04 ± 0.10 | 29.19 ± 0.13 | -0.73 | 0.47 |
| **Time of sampling (days)** | 16.64 ± 1.11 | 19.19 ± 1.03 | -1.65 | 0.10 |
| **Body mass at sampling (g)** | 1643 ± 34 | 1559 ± 28 | -1.87 | 0.06 |
| **Body condition index** | 9.33 ± 0.09 | 8.28 ± 0.09 | -6.12 | <0.0001 |
| **NO1 (μM)** | 112.45 ± 13.29 | 101.42 ± 8.72 | -0.56 | 0.58 |
| **NO2 (μM)** | 192.47 ± 17.79 | 166.31 ± 7.69 | -0.96 | 0.34 |
| **Innate immune response (%)** | 93.34 ± 15.17 | 86.06 ± 14.68 | -0.09 | 0.92 |
| **Corticosterone (ng.ml$^{-1}$)** | 20.17 ± 1.63 | 36.91 ± 8.79 | -2.50 | 0.01 |